\documentclass{aastex631}

\usepackage{xcolor}
\def \myulx {NGC~5907~ULX-1}
\def \ulx {ULX1}
\def \iasfmi {\affiliation{INAF, Istituto di Astrofisica Spaziale e Fisica Cosmica,
  Via Alfonso Corti 12, I-20133, Milano, Italy}}
\def \oar {\affiliation{INAF, Osservatorio Astronomico di Roma, 
  Via Frascati 33, I-00078, Monteporzio Catone, Italy}}
\def \iasfpa {\affiliation{INAF, Istituto di Astrofisica Spaziale e Fisica Cosmica,
  Via Ugo La Malfa 153, I-90146, Palermo, Italy}}
\def \oab {\affiliation{INAF, Osservatorio Astronomico di Brera, 
  Via Brera 28, I-20121, Milano, Italy}}
\def \iuss {\affiliation{Scuola Universitaria Superiore IUSS Pavia, 
  Palazzo del Broletto, Piazza Della Vittoria 15, I-27100, Pavia, Italy}}
\def \comillas {\affiliation{Universidad Pontificia Comillas Madrid-ICAI, 
  Calle de Alberto Aguilera 25, 28015, Madrid, Spain}}

\graphicspath{{./}{figures/}}

\received{22 12 2023}
\revised{23 02 2024}
\accepted{07 03 2024}
\submitjournal{ApJ}

\shorttitle{NGC5907 ULX1 Orbit}
\shortauthors{Belfiore et al.}

\begin{document}

\title{The Orbit of NGC 5907 ULX-1}

\correspondingauthor{Andrea Belfiore}
\email{andrea.belfiore@inaf.it}

\author[0000-0002-2526-1309]{Andrea Belfiore} \iasfmi
\author[0000-0002-9393-8078]{Ruben Salvaterra} \iasfmi
\author[0000-0002-9705-2883]{Lara Sidoli} \iasfmi
\author[0000-0001-5480-6438]{Gian Luca Israel} \oar
\author[0000-0002-0018-1687]{Luigi Stella} \oar
\author[0000-0001-6739-687X]{Andrea De Luca} \iasfmi
\author[0000-0003-3259-7801]{Sandro Mereghetti} \iasfmi
\author[0000-0003-4849-5092]{Paolo Esposito} \iuss \iasfmi
\author[0000-0002-3869-2925]{Fabio Pintore} \iasfpa
\author[0000-0002-5042-1036]{Antonino D'A\`i} \iasfpa
\author[0000-0003-3952-7291]{Guillermo Rodr\`iguez Castillo} \iasfpa
\author[0000-0001-5819-3552]{Dominic J. Walton}
  \affiliation{Centre for Astrophysics Research, University of Hertfordshire, College Lane, Hatfield AL10 9AB, UK}
  \affiliation{Institute of Astronomy, University of Cambridge, Madingley Road, Cambridge CB3 0HA, UK}
\author[0000-0003-0388-0560]{Felix F\"urst}
  \affiliation{European Space Agency (ESA), European Space Astronomy Centre (ESAC),
  Camino Bajo del Castillo s/n, 28692 Villanueva de la Ca\~{n}ada, Madrid, Spain}
\author[0000-0001-6337-2184]{Danilo Magistrali} \comillas
\author[0000-0001-5480-9835]{Anna Wolter} \oab
\author[0000-0001-8688-9784]{Matteo Imbrogno}
  \affiliation{Dipartimento di Fisica,
  Universit\`a degli Studi di Roma “Tor Vergata”, 
  via della Ricerca Scientifica 1, I-00133 Rome, Italy}
  \oar
  \affiliation{Universit\`a degli Studi di Roma “La Sapienza”,
  Piazzale Aldo Moro 5, I-00185 Rome, Italy}

\begin{abstract}

We report on the orbit of the binary system powering the most extreme
ultraluminous X-ray pulsar known to date: \myulx~(hereafter \ulx).
ULX1 has been the target of a substantial multi-instrument campaign, mainly in
the X-ray band, but no clear counterparts are known in other bands.
Although \ulx\ is highly variable and pulsations can be 
transient (regardless of the source flux),
the timing data collected so far allow us
to investigate the orbit of this system.
We find an orbital period $P_{orb}=5.7^{+0.1}_{-0.6}\text{\,d}$
and a projected semi-axis $A_1 =3.1^{+0.8}_{-0.9}\text{\,lts}$.
The most likely ephemeris is:  $P_{orb}=5.6585(6)\text{\,d}$,
$A_1 = 3.1(4)\text{\,lts}$, and the epoch of ascending nodes passage is:
$T_{asc} = 57751.37(5)\text{\,MJD}$.
However, there are 6 similar solutions, acceptable within $3\,\sigma$.
We find further indications that \ulx\ is a high-mass X-ray binary.
This implies that we are observing its orbit face-on, with an inclination $<5\text{\,deg}$.
\end{abstract}


\section{Introduction} \label{sec:intro}

\myulx~(\ulx) is the most luminous member of the known ultraluminous X-ray pulsars (PULXs), peaking at an apparent luminosity of $L_{\rm{X,peak}} \sim 10^{41}\,\text{erg s}^{-1}$.
PULXs are an emerging class of accreting X-ray pulsars with luminosity far in excess
of the Eddington limit for a neutron star.
This is a sub-class of ultraluminous X-ray sources (ULXs), i.e.
X-ray sources, located off-center of their host galaxy, whose isotropic luminosity
is greater than $10^{39}\,\text{erg s}^{-1}$ (\citealt{Walton2022, Tranin2023};
for recent reviews see \citealt{King2023,Pinto2023}).
PULXs are accreting pulsars, likely in high mass X-ray binary (HMXB) systems, 
and thus are neutron stars orbiting a stellar companion.
Their accretion geometry is not spherical and their magnetic field is so strong
that the Eddington limit -- which assumes spherical symmetry and Thomson
cross-section -- does not formally apply.
Still, it remains a useful point of reference for comparison with other X-ray binary
systems.
Given the extreme nature of PULXs, it is important to investigate the nature of their
companions and to measure the orbital parameters of these systems, as they reflect
the conditions under which accretion at such extreme rates can occur.
Because the detection of pulsations depends on several factors (including pulsed
fraction, photon statistics, and background level),
other known ULXs may yet turn out to be PULXs, as we keep observing them \citep{King2016,Pintore2017,Walton2018}.

After the discovery of pulsations with XMM-Newton and NuSTAR \citep{Israel2017a},
\ulx\ has been regularly monitored with the Neil Gehrels Swift Observatory 
and observed on numerous occasions with XMM-Newton, NuSTAR, and Chandra \citep{Fuerst2023}.
The neutron star powering \ulx\ shows strong long-term variability, exhibiting a high state
that can last for years ($L_{\rm{X,peak}} \sim 10^{41}\,\text{erg s}^{-1}$), during which its flux
is modulated over a period of $78\,\text{d}$ \citep{Walton2016},
as well as a low state ($L_{\rm{X}} < 10^{39}\,\text{erg s}^{-1}$) during which a spatially
extended X-ray nebula is revealed (which is otherwise drowned out by the emission from the
point source; \citealt{Belfiore2020}).
The source can transition between these high- and low-flux states within days \citep{Walton2015}.
The pulsed fraction of \ulx\ seems to vary randomly across different observations and its spin
period $P_{spin}$ evolves noticeably, driven by the strong torque that results from the accretion of 
matter at a very high rate (which is also responsible for its extreme luminosity;
\citealt{Fuerst2023}).
The earliest detection of pulsations revealed a spin period of 1.43\,s in 2003,
with the neutron star having subsequently been spun up to spin periods of 1.14\,s in 2014 
and 0.95\,s in 2017.
This erratic behavior hampers timing studies of the pulsar.

Because the host galaxy, NGC~5907, is nearly edge-on, our line of sight to \ulx\  
is heavily obscured by dust. Optical/NIR searches for its counterpart have thus proven difficult.
We are therefore forced to rely only on X-ray timing of the pulsar to infer
the orbital parameters of this system.
A first estimate of the period ($P_{orb}$), projected semiaxis 
($A_1=a_{ns} \cdot \text{sin}i$), and epoch of ascending nodes ($T_{asc}$)
of the orbit of the neutron star came together with the discovery of 
pulsations \citep{Israel2017a}.
Such an analysis was based on two NuSTAR observations taken in July 2014, with a
baseline of 4.7 days.
They report, at $1~\sigma$ confidence,
$P_{orb}=5.3^{+2.0}_{-0.9}\,\text{d}$ and
$A_1=2.5^{+4.3}_{-0.8}\,\text{lts}$.
However, at $3~\sigma$ confidence, only lower limits on $A_1>1.4\,\text{lts}$ and
$P_{orb}>4.0\,\text{d}$ were obtained, whereas upper limits rely on physical
considerations about the mass of the companion.

In this paper we consider all of the observations taken so far with XMM-Newton
and NuSTAR and derive an updated orbital ephemeris.
Sec.\,\ref{sec:obsprep} describes the data used for this paper and how
they were selected. Sec.\,\ref{sec:analysis} describes the timing analysis
that leads to our results, presented in Sec.\,\ref{sec:results}. 
A discussion follows in Sec.\,\ref{sec:discussion}, and conclusions are
drawn in Sec.\,\ref{sec:conclusion}.

\section{Observations and Data Preparation} \label{sec:obsprep}

Timing \ulx\ requires sufficient photon statistics and good time resolution, which restricts our analysis to the data obtained by two X-ray observatories:
NuSTAR \citep{Harrison2013} and XMM-Newton \citep{Jansen2001}.
The rapid and erratic spin evolution of \ulx\ does not allow for coherent
timing on long timescales (more than a few weeks), while on short timescales
the intrinsic behaviour of the pulsar (e.g. the accretion-driven
spin-up) can account for any linear trend
in spin period.
Therefore we must rely on clusters of 2 or more observations, all taken
within a couple of weeks of each other, during which pulsations are detected.
Any non-linearity in the spin evolution within each observation cluster can be
ascribed, to a first approximation, to the orbit of the system.

So far, 3 clusters of observations that meet the above requirements are available (see table\,\ref{tab:obstable}):
3 observations in 2014 (cluster A; these are the data that led to the initial discovery of
pulsations and to the first orbital ephemeris),
2 observations in 2017 (cluster B), and
3 observations in 2019 (cluster C).

\begin{table}
  \begin{center}
    \begin{tabular}{|c|c|c|c|c|c|}
\hline 
cluster & obs. ID & observatory & date & duration (ks) & photons \tabularnewline
\hline 
\hline 
A & 0729561301 & XMM-Newton & 2014-07-09 & 42 (42) & 12879\tabularnewline
\hline 
A & 80001042002 & NuSTAR & 2014-07-09 & 57 & 3297\tabularnewline
\hline 
A & 80001042004 & NuSTAR & 2014-07-12 & 56 & 3291\tabularnewline
\hline 
B & 0804090301 & XMM-Newton & 2017-07-02 & 40 (32) & 3394\tabularnewline
\hline 
B & 0804090401 & XMM-Newton & 2017-07-04 & 36 (36) & 2221\tabularnewline
\hline 
C & 0824320201 & XMM-Newton & 2019-06-12 & 60 (59) & 12216\tabularnewline
\hline 
C & 0824320301 & XMM-Newton & 2019-06-19 & 49 (49) & 9069\tabularnewline
\hline 
C & 0824320401 & XMM-Newton & 2019-06-26 & 64 (54) & 7854\tabularnewline
\hline 
    \end{tabular}
  \end{center}
  \caption{
    X-ray observations used in our analysis. 
    For XMM-Newton, in parentheses is the net exposure time, after
    the removal of high background periods.
    The number of photons is measured after all filters have been applied.
    The baselines for the 3 clusters are: $404\,\text{ks}$ for cluster A, 
    $222\,\text{ks}$ for cluster B, and $1271\,\text{ks}$ for cluster C.
  }
  \label{tab:obstable}
\end{table}

The XMM-Newton observations were taken with the the EPIC-PN camera
\citep{Struder2001} in Full
Frame mode, and thus have a time resolution of 73.4 ms.
We do not consider the EPIC-MOS data because its time resolution is
not sufficient for the timing analysis of a $\sim1\,\text{s}$ pulsar.
We used the XMM-Newton Scientific Analysis System \citep[SAS v21.0][]{Gabriel2004} to reprocess and
filter the events, and to correct their time of arrival to the Solar
System barycenter (using the DE200 ephemeris).
We adopt the position obtained for \ulx\ with Chandra \citep{Sutton2013}:
\mbox{R.A. = 15h~15m~58.62s~$\pm$~0.01s},
\mbox{Dec = +56$^{\circ}$~18'~10.3"~$\pm$~0.1"} (J2000).
We applied standard quality filters and excluded periods of high
background, as recommended by the XMM-Newton team.
We kept all the events within 30$''$ of the position of \ulx\
with energies $E>1\,\text{keV}$.
These criteria maximise the strength of the pulsed signal for this
particular pulsar \citep{Israel2017a}.

The NuSTAR observations used data from both focal plane modules (FPMA and FPMB)
which have a time resolution of 2\,$\mu$s.
They were reduced with the NuSTAR Data Analysis Software (NuSTARDAS v2.1.2).
We applied the standard quality filters recommended by the NuSTAR team, and kept
all the events within 49$''$ of the position of \ulx\ with energies in the 
$3-15\,\text{keV}$ range.

We again shifted the time of arrival of the photons to the Solar System barycenter.

\section{Data Analysis} \label{sec:analysis}

The pulse profile of \ulx\ is well approximated by a sinusoid.
We construct a model for the evolution of the period of this sinusoid
and fit it directly to the time of arrival of each photon, using
an unbinned likelihood analysis \citep[suppl.\,mat.]{Israel2017a}.
The most likely set of parameters in our model is our best-fit solution.
We then perturb the optimal solution by varying each single parameter.
As we shift one parameter we profile the likelihood by maximising it
over all the other parameters.
We then estimate the uncertainties by measuring the drop in likelihood and
applying Wilks' theorem \citep{Cowan2011}.

Our model accounts for an evolution of the intrinsic spin period
of the pulsar (due to accretion or other torques) and
the Doppler modulation induced by the orbital motion.
We assume that within each cluster of observations (taken less than
2 weeks apart) the intrinsic evolution of the spin period $P$ is linear,
i.e. its time derivative $\dot{P}$ is constant.
We do not assume any relation between the spin parameters taken
in different clusters of observations as the accretion
rate is variable and hardly predictable.
Any non-linearity in the spin evolution observed within
a cluster of observations is ascribed to the orbital modulation.
We assume that the orbit is circular and that its parameters
(the projected semi-axis $A_1$, the orbital period $P_{orb}$
and the epoch of ascending node passage $T_{asc}$) do not
change across different clusters.

Our assumption of a circular orbit is not granted, a priori.
However, we can take it as a first order approximation.
In particular, because most often the pulsar is far from periapsis,
we expect a very limited bias due to this assumption.
As more timing data become available, this model can be extended
to account for an eccentric orbit.

It is clear that the secular spin evolution is intrinsic, because
no orbit (not even around a supermassive black hole) could account
for a change in the spin period of $>10$\%, as observed for
\ulx\ \citep{Fuerst2023}.
However, on short timescales the shift in $\dot{P}$ induced by a binary orbit can be
of the same order as the intrinsic $\dot{P}$.
Therefore this observable is fully degenerate with the unknown
intrinsic spin-up (or spin-down) and we cannot build upon that
to constrain the orbit.

We start by considering each cluster of observations by itself.
This analysis provides weak independent constraints on the 3 orbital
parameters.
In order to minimize the correlation between $P_{orb}$ and $T_{asc}$,
we keep $T_{asc}$ as close as possible to the midpoint of each cluster.
We focus on $A_1$ and $P_{orb}$, by comparing their estimates in each
set of observations and combining them.
The analysis of each cluster maximises the likelihood over all the other parameters:
$P$, $\dot{P}$ and the orbital phase $T_{asc}$.

Afterwards, we shift our estimates of $T_{asc}$ to a common epoch
(close to the midpoint of all data) by adding or subtracting an
integer number of full orbits.
Generally, given a pair $A_1$ and $P_{orb}$, the values of $T_{asc}$
shifted from clusters A and C do not match. 
Forcing them to be the same, while considering both data sets at once,
introduces some aliases in $P_{orb}$ separated by:
\begin{equation}
    \Delta P_{orb}\simeq\frac{P_{orb}^{2}}{T_{C}-T_{A}}\simeq0.018\text{\,d}
\end{equation}
Finally, we examine all these aliases, considering data from all
3 clusters at the same time.

\section{Results} \label{sec:results}

\begin{figure}
  \begin{centering}
    \includegraphics[width=9cm]{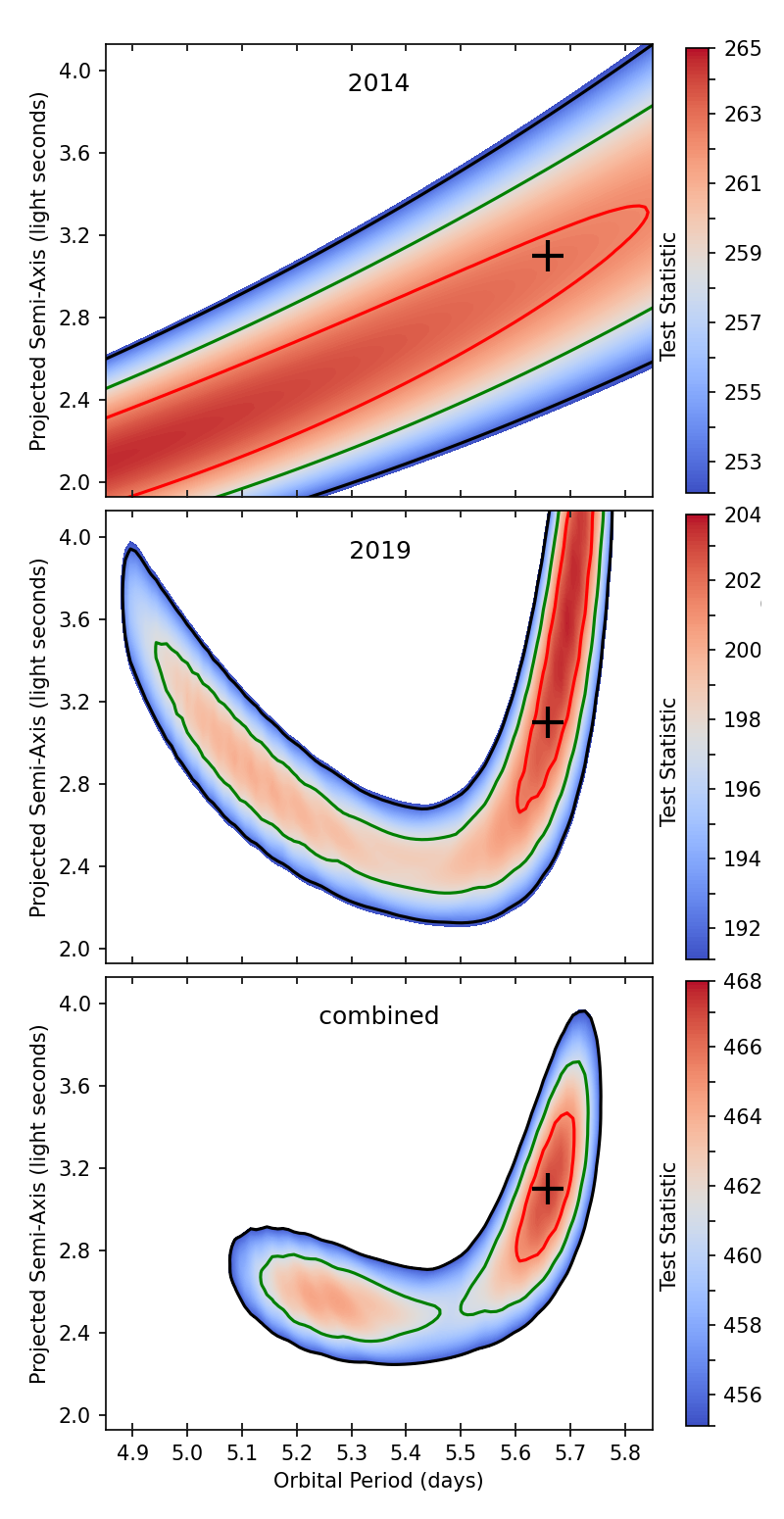}
  \end{centering}
  \caption{
    Estimates of the orbital period ($P_{orb}$, on the X-axis) and
    the projected semi-axis ($A_1$, on the Y-axis) of the orbit of \ulx.
    These estimates are profiled over the orbital phase ($T_{asc}$).
    The upper panel comes from the observation cluster A, the central
    panel from the observation cluster C, and the bottom panel combines
    the above two estimates (without fixing $T_{asc}$ between them).
    Some aliasing is visible in the central panel (propagated to
    the bottom panel).
    A black cross marks the most likely value of $P_{orb}=5.66\text{\,d}$
    and $A_1=3.10\text{\,lts}$, from the combined estimate shown
    in the bottom panel.
    Level curves indicate the 1-$\sigma$ (red), 2-$\sigma$ (green),
    and 3-$\sigma$ (black) contours, for 2 degrees of freedom.
  }
  \label{fig:uncoherent}
\end{figure}

The likelihood analysis within each cluster of observations
provides non-linear constraints on the orbital parameters.
In particular, only clusters A and C provide independent estimates
of all the 3 orbital parameters.
The analysis of cluster B, which has a much shorter baseline,
provides looser constraints, in which each of the orbital
parameters is fully degenerate with the other two.
Therefore, we used only the analysis of clusters A and C to
derive a first estimate for $A_1$ and $P_{orb}$
(see fig.\,\ref{fig:uncoherent}).

We then combined the two estimates of $A_1$ and $P_{orb}$, without enforcing
coherence in the orbital phase between the two epochs.
The most likely values are $P_{orb}=5.66\text{\,d}$ and 
$A_1=3.10\text{\,lts}$ and, at $3\,\sigma$, 
$5.0\text{\,d}<P_{orb}<5.8\text{\,d}$ and
$2.3\text{\,lts}<A_1<4.0\text{\,lts}$.
These values are consistent to within $1\,\sigma$ with the individual estimates obtained for each of the clusters when considered independently (again see fig.\,\ref{fig:uncoherent}).
Therefore, our approximation with a circular orbit seems to be justified.

Finally, we fixed the orbital phase between the two main clusters, A and C, and also incorporated the data from cluster B.
As described above, in sec.\,\ref{sec:analysis}, forcing the coherence
in orbital phase between clusters A and C induces aliases in $P_{orb}$.
Many of these aliases can be ruled out as they imply a value of $T_{asc}$
which is not consistent with the observations in cluster B.
Only 7 aliases are acceptable to within $3\,\sigma$, and one of them
stands out (ID 320, corresponding to 320 full orbits between $T_A$
and $T_C$, see table\,\ref{tab:aliases}).

\begin{table}
  \begin{center}
    \begin{tabular}{|c|c|c|c|c|c|c|c|c|}
\hline 
ID & TS & $N_{\sigma}$ & $R_A$ & $R_B$ & $R_C$ & $P_{orb}$ [d] & $A_1$ [lts] & $T_{asc}$ [MJD] \tabularnewline
\hline
\hline 
320 & 543.83 & - & 262.76 & 76.53 & 202.92 & 5.6585(6) & 3.1(4) & 57751.37(5) \tabularnewline
\hline 
343 & 541.53 & 1.5 & 263.80 & 76.78 & 199.32 & 5.2753(3) & 2.5(3) & 57753.65(5) \tabularnewline
\hline 
338 & 540.11 & 1.9 & 263.19 & 76.79 & 198.51 & 5.3538(3) & 2.5(3) & 57751.05(5) \tabularnewline
\hline 
348 & 539.41 & 2.1 & 263.13 & 75.55 & 199.12 & 5.1992(2) & 2.6(2) & 57750.97(4) \tabularnewline
\hline 
325 & 539.36 & 2.1 & 262.09 & 75.98 & 199.68 & 5.5702(4) & 2.7(3) & 57754.06(4) \tabularnewline
\hline 
333 & 536.21 & 2.7 & 262.00 & 74.71 & 197.90 & 5.4348(2) & 2.5(1) & 57753.83(3) \tabularnewline
\hline 
351 & 535.05 & 2.9 & 261.88 & 73.50 & 198.10 & 5.1546(1) & 2.7(1) & 57753.52(1) \tabularnewline
\hline 
    \end{tabular}
  \end{center}
  \caption{
    Coherent orbital solutions acceptable at the 3-$\sigma$ level
    in a single parameter.
    The ID corresponds to the number of full orbits between the observations
    in clusters A and C.
    The drop in test statistic (TS) from the optimum solution asymptotically follows a $\chi ^2$ distribution
    with 1 degree of freedom, and is converted to $\sigma$ units (i.e. $N_{\sigma}$).
    We report also the Rayleigh TS (a measure of the strength of the signal)
    for each cluster of observations ($R_A$, $R_B$, and $R_C$).
    The orbital parameters are: the orbital period $P_{orb}$, the projected
    semi-axis $A_1$, and the epoch of ascending nodes $T_{asc}$.
    All the uncertainties (in parentheses after the last digit)
    are at $3\,\sigma$ on a single parameter,
    with respect to the most likely solution (ID 320).}
\label{tab:aliases}
\end{table}

\begin{figure}
  \begin{centering}
    \includegraphics[width=10cm]{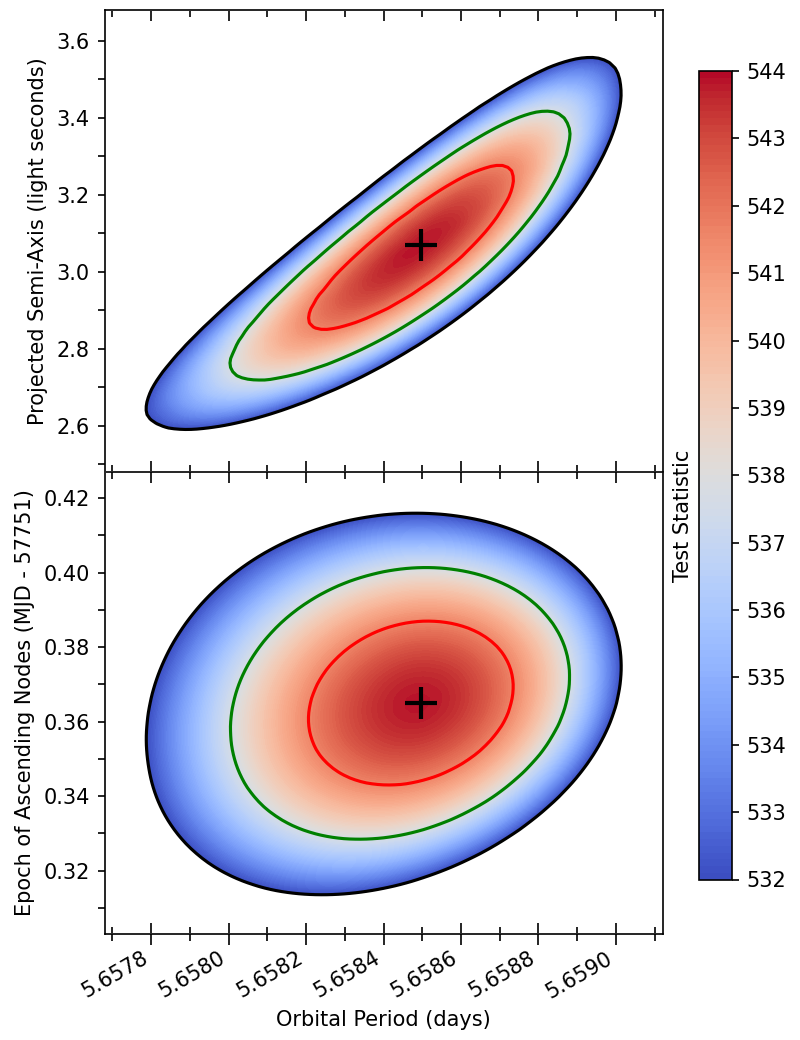}
  \end{centering}
  \caption{
    Estimates of the orbital parameters for the most likely alias
    (ID 320 in table\,\ref{tab:aliases}) in the orbital period ($P_{orb}$,
    on the X-axis) for \ulx\ based on our final, combined analysis.
    The upper panel shows the projected semi-axis ($A_1$) on the Y-axis.
    The lower panel shows the epoch of passage of the ascending nodes
    ($T_{asc}$) on the Y-axis.
    The black cross marks the most likely orbital ephemeris.
    Level curves indicate the 1-$\sigma$ (red), 2-$\sigma$ (green),
    and 3-$\sigma$ (black) contours, for 2 degrees of freedom.
}
  \label{fig:coherent}
\end{figure}

The most likely value of the orbital parameters, for alias 320,
is: $P_{orb}=5.6585(6)\text{\,d}$, $A_1=3.1(4)\text{\,lts}$,
and $T_{asc}=\text{MJD\,}57751.37(5)$.
The reported uncertainty, at $3\,\sigma$ on a single parameter of interest,
is in parentheses after the last digit.
As noted previously, $T_{asc}$ is taken close to the midpoint of the observing baseline,
minimising the correlation between $P_{orb}$ and $T_{asc}$.
The correlation between $P_{orb}$ and $A_1$ does not depend on our
choice of $T_{asc}$ (see fig.\,\ref{fig:coherent}).

\begin{figure}
  \begin{centering}
    \includegraphics[width=15cm]{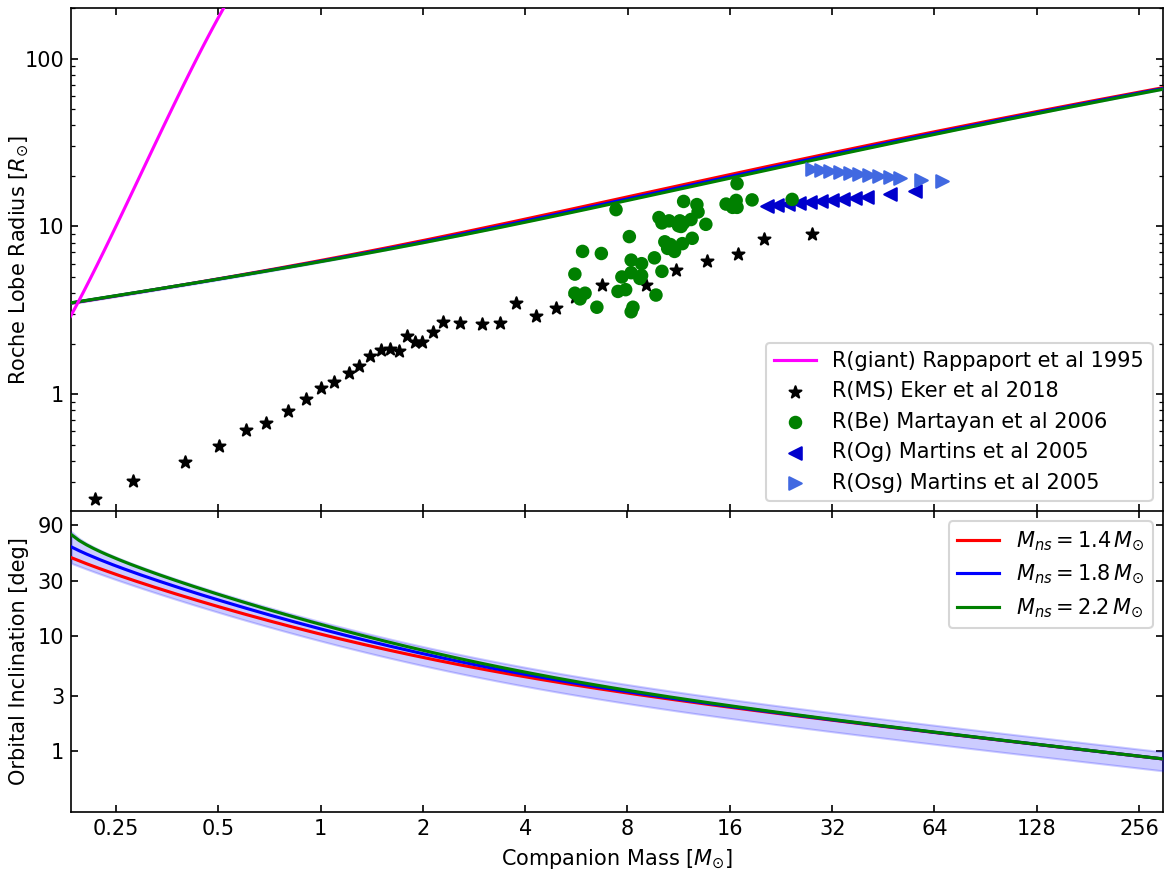}
  \end{centering}
  \caption{
    Geometrical parameters of the \ulx\ system for 3 values of the
    mass of the neutron star ($M_{ns}=1.4,\,1.8,\,2.2\,M_{\odot}$,
    red, blue and green lines, respectively) depending on the mass
    of the companion, $M_c$, on the X-axis.
    Shaded regions cover the 3-$\sigma$ uncertainty ranges on the orbital
    parameters, assuming $M_{ns}=1.8\,M_{\odot}$.
    \textbf{Upper panel}:
    on the Y-axis is the Roche lobe radius $R_l$ in $R_{\odot}$ units
    (see eq.\,\ref{eq:Rl}).
    Stellar radii ($R_{star}$) are also reported for different masses
    and stellar types for comparison
    (data from \cite{Rappaport1995,Eker2018,Martayan2006,Martins2005}).
    For mass transfer to occur $R_{star} \simeq R_l$.
    This rules out a main-sequence companion.
    \textbf{Lower panel}:
    on the Y-axis is the system inclination (see eq.\,\ref{eq:sinI}).
    Unless $M_c$ is very low, the orbit is $\sim$face-on.
  }
  \label{fig:inclination}
\end{figure}

From these constraints, we can derive some other
parameters of the system and its geometry: the Roche lobe
radius and the inclination of the orbital plane.
These parameters depend on the mass of the neutron star,
$M_{ns}$ and the mass of its companion $M_c = q \cdot M_{ns}$ (where $q$ is the mass ratio of the binary components).
We take 3 representative values of $M_{ns}=1.4,\,1.8,\,2.2\,M_{\odot}$
and plot these quantities for a range of $M_c$ (see
fig.\,\ref{fig:inclination}).
We note that the dependence of our results on $M_{ns}$ is very
weak.
For $M_{ns}=1.8\,M_{\odot}$, we indicate with a shaded band
the uncertainty, at $3\,\sigma$, considering any of the 7
acceptable aliases in $P_{orb}$.

We estimate the Roche lobe radius $R_l$ from the semi-major
axis of the orbit, $a$, obtained from $P_{orb}$ through Kepler's
third law:
\begin{equation}
  R_l = f\left(q\right) \cdot
      \left( \frac{GM_{ns}}{4\pi^2} \right)^{\frac{1}{3}}
      \left( 1 + q \right)^{\frac{1}{3}} P_{orb}^{\frac{2}{3}}
    \simeq 2.062
      \frac{q^{\frac{2}{3}} \left( 1 + q \right)^{\frac{1}{3}}}
        {0.6 q^{\frac{2}{3}} + \ln \left( 1 + q^{\frac{1}{3}}\right)}
      \left( \frac{M_{ns}}{M_{\odot}} \right)^{\frac{1}{3}}
      \left( \frac{P_{orb}}{\text{d}} \right)^{\frac{2}{3}}
      \,R_{\odot}
  \label{eq:Rl}
\end{equation}
where we adopted the approximation
$R_l \simeq f\left(q\right) \cdot a$ \citep{Eggleton1983}.
The upper panel of fig.\,\ref{fig:inclination} shows the relation
$R_l\left(M_c\right)$ for our estimate of $P_{orb}$.

We infer the inclination $i$ of the orbital plane with
respect to the line of sight by comparing $a$, obtained from
$P_{orb}$, and the observed $A_1 = \frac{q}{1+q} a \sin i$:
\begin{equation}
\sin i = \left( \frac{GM_{ns}}{4\pi^2} \right)^{-\frac{1}{3}}
      \frac{\left( 1 + q \right)^{\frac{2}{3}}}{q}
      P_{orb}^{-\frac{2}{3}} A_{1}
    = 0.102 \frac{\left( 1 + q \right)^{\frac{2}{3}}}{q} 
      \left( \frac{M_{ns}}{M_{\odot}} \right)^{-\frac{1}{3}}
      \left( \frac{P_{orb}}{\text{d}} \right)^{-\frac{2}{3}}
      \left( \frac{A_1}{\text{lts}} \right)
  \label{eq:sinI}
\end{equation}
The lower panel of fig.\,\ref{fig:inclination} shows the relation
$i\left(M_c\right)$ for our estimate of $P_{orb}$ and $A_1$.
The lack of eclipses or dips in the light curve of \ulx,
instead observed for M51\,ULX7 \citep{Vasilopoulos2021,Hu2021},
implies that $\sin i < \sqrt{1 - f^2\left(q\right)}$,
hence $i<78\,\text{deg}$.
If $M_{c}>5M_{\odot}$, then the orbit must be nearly face-on, with 
$i<5\,\text{deg}$.

\section{Discussion} \label{sec:discussion}

\begin{figure}
  \begin{centering}
    \includegraphics[width=18cm]{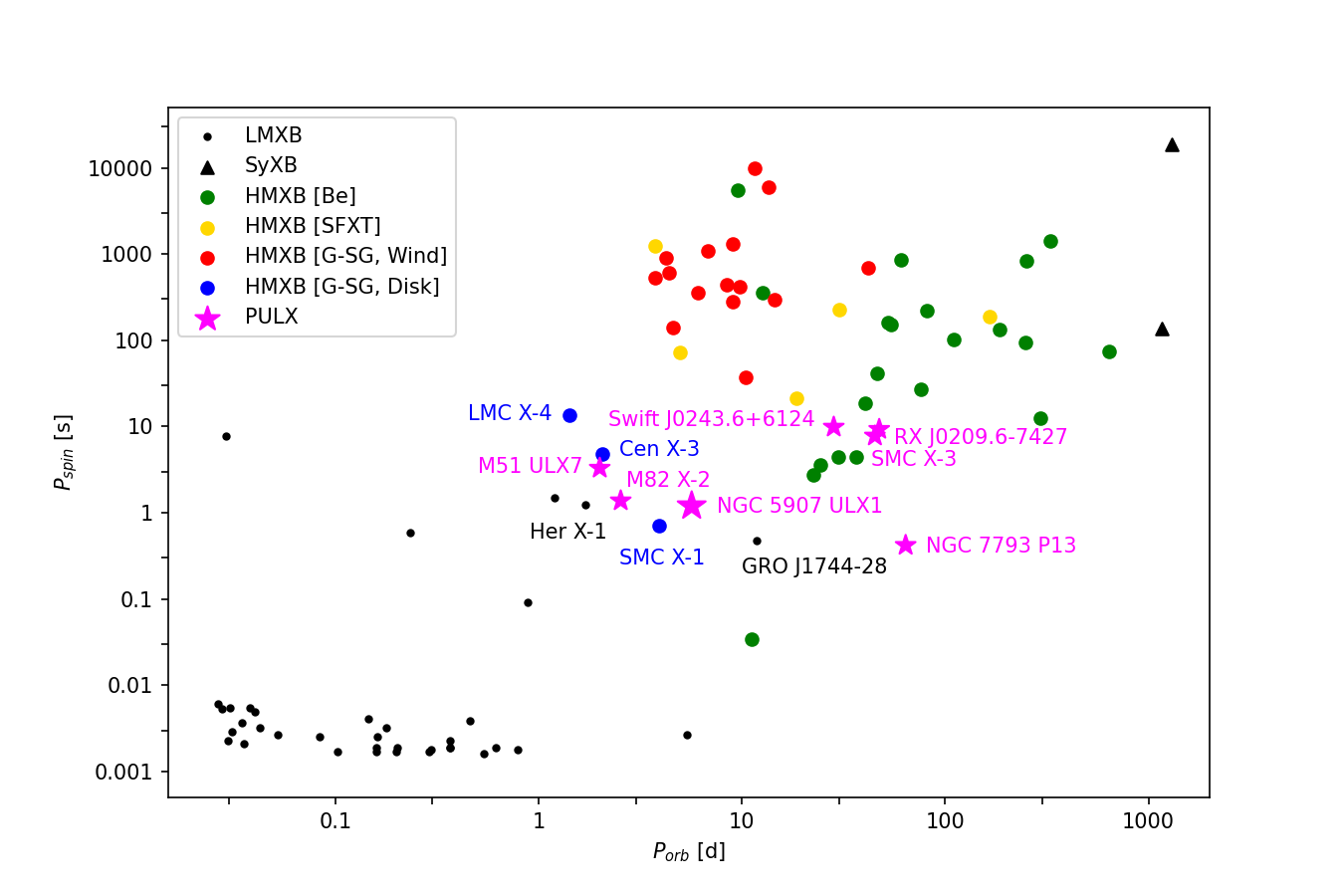}
  \end{centering}
  \caption{
    Distribution of the orbital period ($P_{orb}$, in days, on the 
    X-axis) versus spin period ($P_{spin}$, in seconds, on the Y-axis)
    for known pulsars in X-ray binaries.
    The data shown are from \cite{Fortin2023}, \cite{Neumann2023}, \cite{Avakyan2023},
    \cite{Liu2005} and \cite{Esposito2016}.
    Symbols and colors correspond to the legend, where SyXB
    indicates symbiotic X-ray binaries and SFXT indicates supergiant
    fast X-ray transients.
    We labelled all the PULXs that currently have a firm estimate of $P_{orb}$
    as well as some other systems which lie at a position close
    to that of \ulx\ in this diagram.
  }
  \label{fig:corbet}
\end{figure}

We have undertaken a multi-epoch X-ray analysis of the ULX pulsar \myulx (\ulx)
in order to place updated constraints on its orbital parameters 
by means of X-ray timing.
This information is key to understanding the nature of this remarkable binary system,
and can only be accessed via such studies in the X-ray band given that the distance
to the source and the level of obscuration towards it have prevented the detection
of any stellar counterpart at other wavelengths (e.g. \citealt{Heida2019}).
Using our updated constraints for \ulx, we start by comparing its orbital 
and spin period ($P_{orb}$ and $P_{spin}$, respectively) with those of other
X-ray binaries containing a pulsar \citep{Corbet1984}, see fig.\,\ref{fig:corbet}.
We also included in the plot the other PULXs with a firm estimate
of their orbital period: M82\,X-2 \citep{Bachetti2014};
M51\,ULX-7 \citep{Rodriguez2020};
NGC\,7793\,P13 \citep{Israel2017b,Fuerst2016,Fuerst2021};
Swift\,J0243.6+6124 \citep{Wilson2018,Tsygankov2018}; 
RX\,J0209.6-7427 \citep{Vasilopoulos2020,Chandra2020,Hou2022};
SMC\,X-3 \citep{Townsend2017,Liu2022}.

Although this plot cannot convey the full complexity of
the behavior of these sources, it captures some traits
that distinguish different classes of objects (color
coded in fig.\,\ref{fig:corbet}).
These are classified by looking at the companion and considering 
a broader set of parameters, like luminosity and variability,
spectral shape/features, orbital shape and spin-up 
\citep[see][for a recent review]{Chaty2022}.
Low-mass X-ray binaries (LMXBs) have donors with mass
$M_d<1\,M_{\odot}$, are mostly found to have compact orbits and the
accretors spin at high rates.
HMXBs have donors with mass
$M_d>5\,M_{\odot}$, their orbits are larger and the accretors
generally have longer spin periods.
PULXs are in the central region of the Corbet diagram, on
the lower end of $P_{spin}$ and $P_{orb}$ of HMXBs.

The PULXs neighboring \ulx\ in the Corbet diagram are
M82\,X-2 and M51\,ULX-7, which are both known to be HMXBs.
M82\,X-2 has a mass function $f(M)>5.2\,M_{\odot}$
\citep{Bachetti2014}.
M51\,ULX-7 has a mass function $f(M)>8\,M_{\odot}$,
\citep{Rodriguez2020} and candidate OB supergiant
(OBsg) counterparts \citep{Earnshaw2016}.
Their variability patterns show some similarity to
\ulx\ although they are fainter.
Other PULXs, like the Galactic Swift\,J0243.6+6124
and RX\,J0209.6-7427 have larger $P_{orb}$ and $P_{spin}$.
Both of them have a Be companion and show a markedly
different variability pattern: they show a burst at
periastron and occasionally this burst can briefly surpass
the limit of $10^{39}\,\text{erg s}^{-1}$ (i.e. they do not
exhibit these extreme luminosities for extended periods, in
contrast to \ulx\ and many of the other ULX pulsars).

Exploring the other pulsar binaries with parameters similar to
\ulx, we find that these are generally interpreted as
HMXBs accreting through a disk.
The formation of a disk demands a compact orbit, and
therefore a short $P_{orb}$ \citep{Tauris2023}.
A disk is required to attain a very large secular spin-up
and to reach a short $P_{spin}$.
This explains their behavior, common also to M82\,X-2
and M51\,ULX-7: persistent high luminosity and strong
spin-up over long periods.

Indeed, the closest source to \ulx\ is SMC\,X-1, a HMXB
thought to be disk-fed, with a spin period
$P_{spin}=0.71\,\text{s}$, an orbital period  
$P_{orb}=3.89\,\text{d}$, and a B0 supergiant companion
\citep{Falanga2015}.
It is variable, its luminosity can reach up to
the Eddington limit and it spins up constantly over
several years \citep{Brumback2022}.
Its pulsations are transient even in a high state,
just like \ulx \citep{Pike2019}.
Two other peculiar sources are also close to \ulx\ in the
Corbet diagram: Her\,X-1, with a $2\,M_{\odot}$
donor, and GRO\,J1744-28, known for its peculiar bursting behavior.
However, they are both accreting below their Eddington limits
and their phenomenology does not match that of \ulx.

Our estimate of $P_{orb}$, the known $P_{spin}\sim 1\,\text{s}$,
and the analogy with similar systems leads us to interpret \ulx\ 
as a HMXB, potentially with an OBsg companion, accreting through a disk onto a neutron star.
This interpretation is consistent with the size of the
Roche lobe inferred in eq.\,\ref{eq:Rl} (see 
fig.\,\ref{fig:inclination}).
A main sequence star would severely underfill its Roche
lobe given the orbital parameters we find for \ulx.
A red giant could fill its Roche lobe if it has a core mass
$>0.195\,M_{\odot}$ \citep{Rappaport1995}.
However, a highly super-Eddington regime (exceeding the Eddington limit by a factor $\sim$30) can only be
sustained if $P_{orb}\simeq 1\,\text{d}$ \citep{Rappaport1997}.
Therefore, unless a much more efficient configuration can be devised,
that is also compatible with a $P_{orb}\simeq 5.7\,\text{d}$, a red giant
companion is ruled out.
OBsg and Be companions to \ulx\ could fill their Roche lobes,
being largely affected by factors such as
rotation, metallicity, and magnetic field.
A stable super-Eddington regime in which mass is transferred on a nuclear timescale
can indeed be sustained for a supergiant companion,
provided that its outer layer has a large metallicity
gradient \citep{Quast2019}.
This is consistent with the timescale required to fill up
a nebula with hot plasma, that can explain the diffuse
emission observed around \ulx\ \citep{Belfiore2020}.

\section{Conclusion} \label{sec:conclusion}

We have used all the available X-ray data on \ulx\ to extract
an updated orbital ephemeris.
We analysed groups of observations clustered in time, initially without
enforcing coherence in the orbital phase between them.
We estimated the orbital period as 
$P_{orb}=5.7^{+0.1}_{-0.6}\text{\,d}$ and the projected
semi-axis as $A_1 =3.1^{+0.8}_{-0.9}\text{\,lts}$.
This improves on the previous estimate by \cite{Israel2017a}
where, at the same confidence level, only lower limits
were given.
Therefore, as already suspected, the $78\,\text{d}$ periodicity
of \ulx\ \citep{Walton2016} is not an orbital modulation.

We subsequently tried to work out a complete ephemeris by enforcing coherence
in the orbital phase between groups of observations.
We found 7 solutions, mutually exclusive, that are compatible with our
measure of $P_{orb}$ and $A_1$ reported above.
The most likely ephemeris is: $P_{orb}=5.6585(6)\text{\,d}$,
$A_1 = 3.1(4)\text{\,lts}$, and epoch of ascending nodes passage
$T_{asc} = 57751.37(5)\text{\,MJD}$.
A specific ephemeris is needed to assign an orbital phase to each 
observation, enabling future phase-resolved analyses of the available data.
To resolve the remaining ambiguity over the precise orbital parameters, new 
and carefully-devised timing observations are needed.
Any new independent ephemeris would provide a measure of the evolution of $P_{orb}$,
as already done for several other HMXBs \citep{Falanga2015} and recently also
claimed for M82\,X-2 \citep{Bachetti2022}.
Because the orbital decay increases with the mass loss rate of the donor
\citep{Quast2019}, temporal baselines comparable to the existing coverage
could be sufficient to detect these changes. 

Based on our updated results for its orbit, we argue that \ulx\ is a HMXB
that contains a neutron star accreting at extreme rates through a disk
from an OBsg donor.
This implies that the orbit is nearly face-on with an inclination
$i<5\text{\,deg}$.

\begin{acknowledgments}
We acknowledge financial support from ASI under ASI/INAF
agreement N.2017-14.H.0.
GLI, PE, FP, ADA, GRC, and AW acknowledge financial support 
from the Italian Ministry for University and Research 
through the PRIN grant 2022Y2T94C (SEAWIND) and the INAF LG 2023 BLOSSOM.
ADA acknowledges funding from the Italian Space Agency,
contract ASI/INAF n. I/004/11/4.
SM acknowledges support from the Italian Ministry for University and Research,
through grant 2017LJ39LM (UNIAM) and from INAF through a Large Program for Fundamental Research 2022.
FP acknowledges financial support from the Italian Ministry for University and
Research through the grant 2023 (OBIWAN).
This research is based on observations obtained with XMM-Newton,
a European Space Agency (ESA) science mission with instruments 
and contributions directly funded by ESA member states and NASA.
This work also made use of data from NuSTAR, 
a mission led by the California Institute of Technology, 
managed by the Jet Propulsion Laboratory, and funded by NASA.
\end{acknowledgments}

\appendix

\section{Orbital Parameters for All the Acceptable Ephemerides} \label{sec:suppl}

\begin{figure}
  \begin{centering}
    \includegraphics[width=18cm]{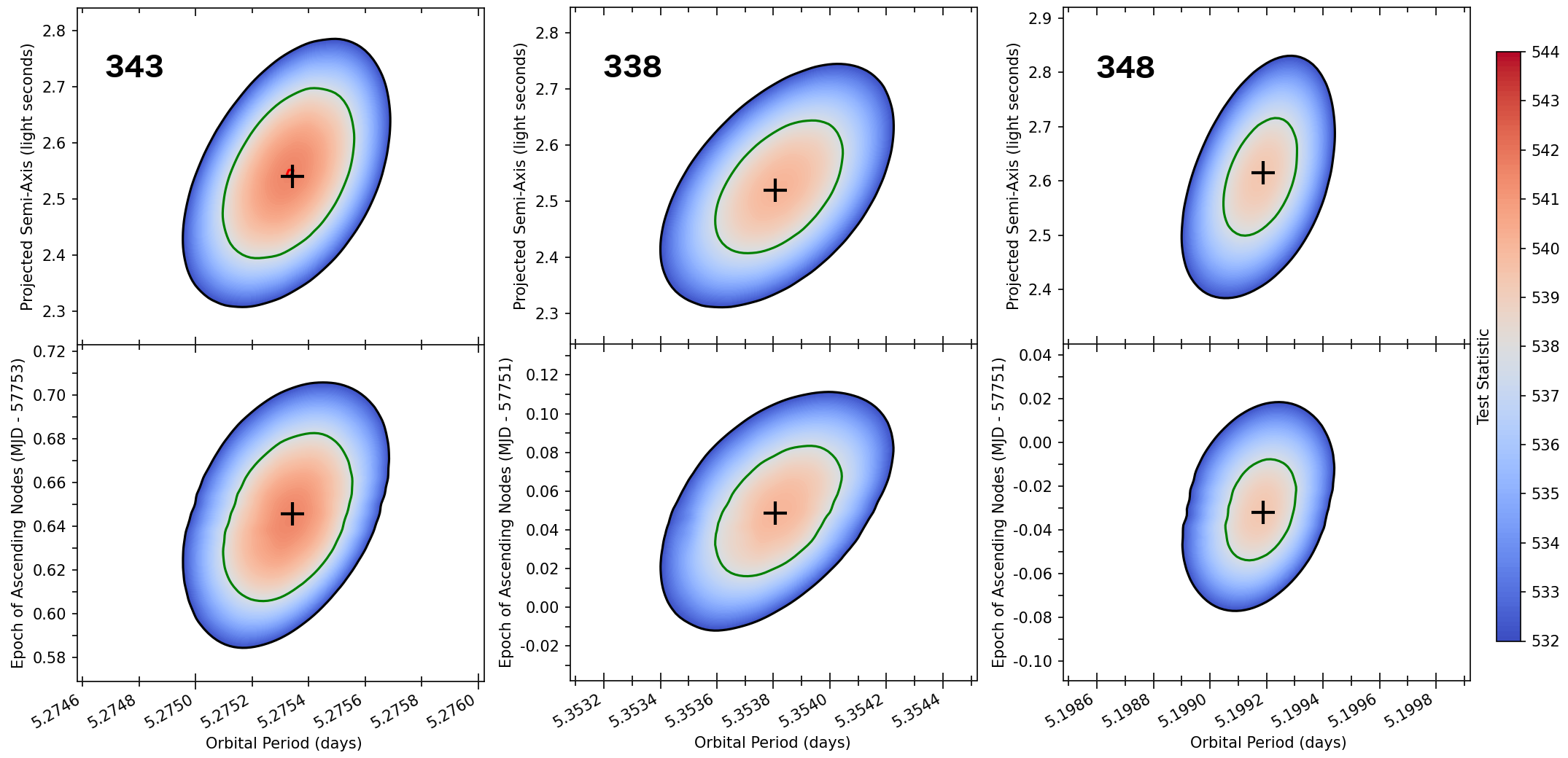}
  \end{centering}
  \caption{
    Estimates of the orbital parameters for the aliases labelled
    with IDs 343, 338, and 348 in table\,\ref{tab:aliases},
    from left to right.
    See fig.\,\ref{fig:coherent} for more details.
  }
  \label{fig:alias1}
\end{figure}

\begin{figure}
  \begin{centering}
    \includegraphics[width=18cm]{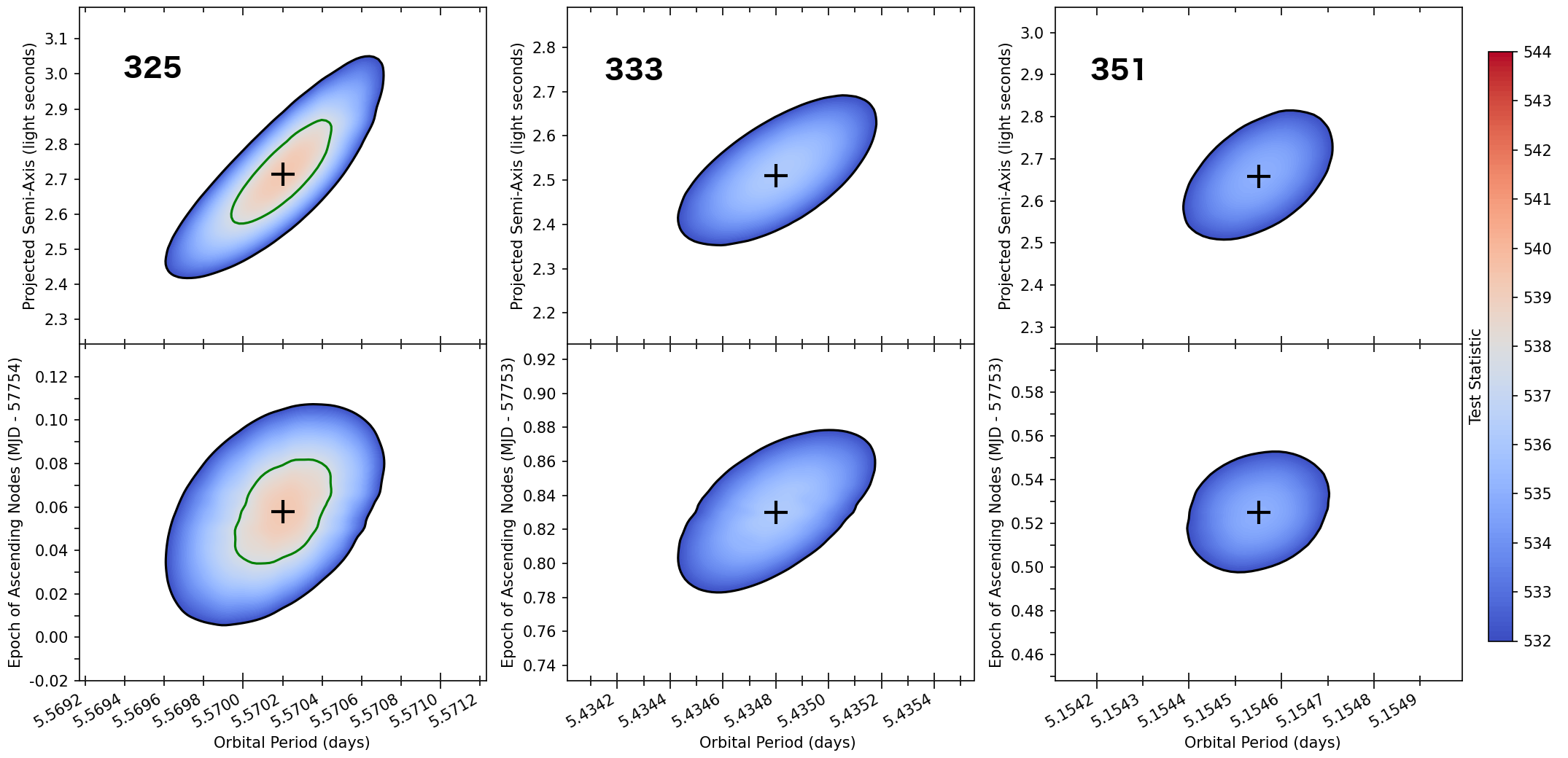}
  \end{centering}
  \caption{
    Estimates of the orbital parameters for the aliases labelled
    with IDs 325, 333, and 351 in table \ref{tab:aliases},
    from left to right.
    See fig.\,\ref{fig:coherent} for more details.
  }
  \label{fig:alias2}
\end{figure}

The main text includes a plot (fig.\,\ref{fig:coherent}) that shows
the correlation between the orbital parameters for the most likely
orbital ephemeris (which we refer to as ID 320).
We report here similar plots for all the other ephemerides listed in Table \ref{tab:aliases} which, while less likely, are all formally acceptable within the $3\,\sigma$ level when compared against the best-fit solution (see Figures \ref{fig:alias1} and \ref{fig:alias2}).
All the contour levels refer to the most likely ephemeris, displayed
in fig.\,\ref{fig:coherent}, and therefore share the same colorbar.

\bibliography{refs}{}
\bibliographystyle{aasjournal}

\end{document}